\documentclass[%
  reprint,
  notitlepage,
  superscriptaddress,
  longbibliography
]{revtex4-1}

\usepackage[english]{babel}
\usepackage[utf8]{inputenc}
\usepackage[caption=false]{subfig}
\usepackage{graphicx}
\usepackage{amsmath}
\usepackage{amssymb}
\usepackage{amsfonts}
\usepackage{bm}
\usepackage[
    space-before-unit,
    range-units=repeat,
    detect-weight=true,
    detect-family=true
]{siunitx}
\usepackage{rotating}

\usepackage{color}

\graphicspath{{graphics/}}

\newcommand {\abs}[1]{\left|{#1}\right|}
\newcommand {\txt}[1]{\text{#1}}
\newcommand {\brc}[1]{\left( #1 \right)}
\newcommand {\sqrbrc}[1]{\left[ #1 \right]}

\begin{document}

\title{Observation of the pressure effect in simulations of droplets splashing on a dry surface.}
\date{\today}
\author{A.M.P. Boelens}
\affiliation{Department of Energy Resources Engineering, Stanford University, 397 Panama Mall, Stanford, California 94305,  USA}
\author{A. Latka}
\affiliation{Department of Physics, University of Chicago, 5720 South Ellis Avenue, Chicago, IL 60637, USA}
\author{J.J. de Pablo}
\affiliation{Institute for Molecular Engineering, University of Chicago, 5801 South Ellis Avenue, Chicago, Illinois 60637, USA}
\email{depablo@uchicago.edu}



\begin{abstract}
At atmospheric pressure, a drop of ethanol impacting on a solid surface produces
a splash. Reducing the ambient pressure below its atmospheric value suppresses
this splash. The origin of this so-called pressure effect is not well understood
and this is the first study to present an in-depth comparison between various
theoretical models that aim to predict splashing and simulations. In this work the
pressure effect is explored numerically by resolving the Navier-Stokes equations
at a 3-nm resolution. In addition to reproducing numerous experimental
observations, it is found that different models all provide elements of what is
observed in the simulations. The skating droplet model correctly predicts the
existence and scaling of a gas film under the droplet, the lamella formation
theory is able to correctly predict the scaling of the lamella ejection velocity
as function of the impact velocity for liquids with different viscosity, and
lastly, the dewetting theory's hypothesis of a lift force acting on the liquid
sheet after ejection is consistent with our results.
\end{abstract}

\maketitle


\section{Introduction}

A wide variety of outcomes can occur when a liquid droplet hits a dry solid
surface. Depending on impact velocity, surface tension, viscosity, ambient
pressure, and surface roughness, one can observe deposition, splashing, receding
breakup, or a rebound of the droplet \cite{yarin2006}. The effect of ambient
pressure on the transition between the smooth droplet-deposition and the
splashing regimes is particularly intriguing. While intuition suggests that
pressure should have a stabilizing effect, it has been found that lowering,
instead of increasing the ambient pressure, suppresses splashing \cite{xu2005}.
Despite various attempts to capture this so-called ``pressure effect'' in
numerical simulations \cite{duchemin2011,taura2010,koplik2013}, its origins are
still unknown. This knowledge gap has hindered technological developments, as there
are numerous applications that could benefit from control over the splashing of
droplets, including erosion, coating, cleaning, cooling, high-throughput drug
screening, and fabrication techniques that rely on inkjet printing technology
\cite{josserand2016,price2014}.

When surveying the literature on splashing experiments, two different observations of the effect of the
ambient gas pressure on splashing have been reported: i) the effect of pressure
that can most readily be observed is that, assuming the crown of the splash at atmospheric
pressure reaches a certain height, as pressure gets reduced the height of the
splash decreases \cite{xu2005}, and ii) when observing the splash using Ultrafast
Interference Imaging \cite{driscoll2011}, a thin gas film can be observed under
the liquid sheet that was ejected from the droplet upon impact. As the ambient
pressure is reduced, at a certain threshold pressure, this gas film is
not observed anymore \cite{latka2017}.

These two observations also help to illustrate the challenges associated with
the multiscale nature of splashing. 
Length scales range from the contact line, where
intermolecular forces act at nanometer length scales, to the characteristic
length scale of the droplet itself, typically on the order of millimeters or
larger. Time scales range from microseconds before impact, when deformation of
the droplet leads to formation of the ``central'' air bubble
\cite{chandra1991,driscoll2011,mandre2012}, to milliseconds after impact when,
as the liquid spreads radially in a thin film or lamella, the liquid sheet that
was ejected from this lamella breaks up into many smaller droplets. While past
efforts have tried to identify a single splashing criterion
\cite{stow1981,stevens2014}, all these length and time scales must be resolved
or modeled in order to fully capture the physics of splashing.

In this paper, by fully resolving the Navier-Stokes equations, we reproduce the
pressure effect on splashing and, in doing so, reveal a number of features that
help explain its origins. This is achieved by performing calculations with spatial
and temporal resolutions as high as $3 \si{\nano\m}$ and $0.5 \si{\pico\s}$.
These are length and time scales which are currently inaccessible to
experiments. To obtain such high resolution, the simulations of droplets
splashing on a dry surface presented in this work are restricted to a
two-dimensional axisymmetric geometry.

In addition to reproducing the pressure effect, the simulations describe several
experimentally observed phenomena, including the formation of the central air
bubble \cite{chandra1991,driscoll2011}, the formation and ejection of a liquid
sheet \cite{field1989,driscoll2011}, and the contact line-instability leading to
entrainment of gas bubbles at the liquid/gas interface \cite{driscoll2010}. A
parameter sweep of lamella formation, varying impact velocity, viscosity, and
surface tension, allows us to evaluate three recently proposed theories of
splashing and lamella formation. One of these proposes air entrainment at the
contact line, also known as dewetting \cite{marchand2012}, as well as the
existence of a lift force \cite{riboux2014}, as the mechanisms responsible for
liquid-sheet formation and breakup. Another theory attributes splashing to a
``skating'' motion of the droplet on a thin gas film, and the deflection of
liquid on the impact surface \cite{mandre2012}. The third theory
\cite{mongruel2009} proposes that lamella formation is caused by the interaction
of viscous and inertial length scales. 

The simulations show that upon impact a very thin gas film is present under the droplet.
While at high enough impact velocities this gas film is not stable and tends to
collapse, it is always present at the edge of the spreading droplet. This is
because the contact line moves at high speed along the surface in a ``rolling'' fashion, continuously extending the gas
film at the edge of the droplet. The observed scaling of the height of a this very thin gas film as function of the impact velocity 
is consistent with literature \cite{mandre2012,kolinski2014}. Simulation results
for an increased slip length on the wall for the gas phase
are consistent with the hypothesis that the breakdown of the continuum
assumption introduces a slip length on the wall which is pressure dependent
\cite{sprittles2017}. This provides a possible explanation for the experimental
observation of a threshold pressure for splashing as mentioned above
\cite{latka2017}.

While it is not possible to identify the correct scaling of the lamella ejection velocity as function of the impact velocity from the currently
available simulation data, the lamella formation theory is the only model that
captures the effect of viscosity as found in the simulations. In
addition, evidence is found for the existence of an early-time viscous length
scale for the height of the lamella as proposed in the lamella formation theory \cite{mongruel2009}, and a new surface
tension based length scale is found. Pertaining to the breakup of the liquid
sheet, our results support the concept of a lift force acting on the liquid
sheet, which was proposed as part of the dewetting theory. Our results
disagree with the idea suggested by the skating-droplet theory that it is the
deflection of liquid on the solid surface that causes a drop to splash.

 
\section{Theory \& Method}

To facilitate the tracking of topological changes, a Volume of Fluid approach
\cite{hirt1981} is adopted in this work. The VOF approach evolves around the
definition of a phase parameter $\alpha$ with the following properties:
\begin{equation}
  \alpha
=
  \left\{
  \begin{array}{ll}
  0      & \txt{in gas phase}    \\
  (0, 1) & \txt{on interface}    \\
  1      & \txt{in liquid phase}
  \end{array}
  \right.
\end{equation}
The evolution of $\alpha$ is calculated using the following transport equation:
\begin{equation}
  \frac{\partial \alpha}{\partial t}
+ \nabla \cdot \brc{\alpha \vec{v}}
+ \nabla \cdot \brc{\alpha \brc{1 - \alpha} \vec{v}_{lg}}
=
  0,
\end{equation}
where $\vec{v}$ is the phase averaged velocity, and $\vec{v}_{lg}$ is 
a velocity field suitable to compress the interface. This equation is
equivalent to a material derivative, but rewritten to minimize numerical
diffusion \cite{rusche2002}.

The phase parameter is used to calculate the phase averaged density, $\rho$,
velocity, $\vec{v}$, and viscosity, $\mu$, which are used in the momentum balance:
\begin{equation}
  \frac{\partial \rho \vec{v}}{\partial t}
+ \nabla \cdot \brc{\rho \vec{v} \otimes \vec{v}}
=
- \nabla p
+ \nabla \cdot \brc{\mu \nabla \vec{v}}
+ \rho \vec{g}
- \vec{f},
\end{equation}
and the continuity equation:
\begin{equation}
  \nabla \cdot \vec{v} 
= 
  0.
\end{equation}
In the above equations $t$ is time, $p$ is pressure, $g$ is gravity, $\vec{f}$ is any body force, like
the surface tension force, and $\otimes$ is the dyadic product. To complete the
VOF model, an expression is needed to calculate the surface tension force
$\vec{f}_{\txt{st}}$, and the initial and boundary conditions need to be chosen. The
surface tension force is calculated using the expression \cite{brackbill1992}:
\begin{equation}
  \vec{f}_{\txt{st}}
=
  \sigma_{\txt{st}} \kappa \nabla{\alpha}
\end{equation}
where $\sigma_{\txt{st}}$ is the surface tension coefficient, and $\kappa$ is
the curvature of the interface. 

The computational domain has two different kinds of boundary conditions for each
variable; on the bottom there is the impact wall, and on the sides and top there
are boundary conditions which allow for the in and
out flow of gas. As boundary condition for $\alpha$ on the impact wall the Generalized Navier
Boundary Condition (GNBC) is used \cite{qian2003,gerbeau2009}. With this boundary
condition the dynamic contact angle is allowed to vary freely, but a restoring
line-tension force is applied at the contact line whenever the dynamic angle
deviates from the equilibrium contact angle. This restoring force is an
additional source term in the Navier-Stokes equations, and has the following
form:
\begin{equation}
  \vec{f}_{\txt{lt}}
=
- \frac{\sigma_{\txt{st}}}{h}
  \cos{\theta_{0}}
  \nabla_{\txt{2D}}{\alpha}
\end{equation}
This force is applied on the liquid-gas interface in the first grid cells next
to the wall and balances the surface tension force when the dynamic contact
angle $\theta$ is equal to $\theta_{0}$. In the above equation
$\sigma_{\txt{st}}$ is the surface tension coefficient, $h =  V/A$, is the mesh
height, with $A$ the surface area of the wall in a grid cell, and $V$ its
volume, and $\nabla_{\txt{2D}}{\alpha}$ is the gradient of $\alpha$ on the wall.
More information on the derivation and validation of this boundary condition can
be found in Ref. \citep{boelens2016b}. On the wall the velocity obeys the
Navier-slip boundary condition with a slip length of $\lambda_{\txt{N}} = 1
\si{\nano\m}$. While the slip length for the gas phase should be larger than
this value, a value of one nanometer is realistic for a contact line. For mass
conservation, the boundary condition for the pressure enforces both a zero flux
and a zero second derivative of the pressure normal to the wall. 

On the other sides of the simulation box the phase parameter, $\alpha$, obeys a
Dirichlet fixed-value boundary condition for inflow and a zero-gradient Neumann
boundary condition for outflow. The Dirichlet fixed value is set to zero, which
is equivalent to only allowing gas to flow in. For each grid cell next to a side
wall the local inflow velocity is calculated from the wall-face normal component
of the velocity vector associated with the center of that specific grid cell.
The boundary condition for the outflow velocity is a zero-gradient Neumann
boundary condition. The total pressure $p_{0}$ on the side walls is kept
constant according to a simplified Bernoulli's equation:
\begin{equation}
p_{0}
=
p + \frac{1}{2} \rho \abs{\vec{v}}^{2} \;.
\end{equation}
In this boundary condition the gas is assumed inviscid, and when the gas inflow
velocity, $\vec{v}$, changes, the pressure, $p$, changes accordingly. The above
equations are solved using the VOF solver of the OpenFOAM Finite Volume toolbox
\cite{openfoam}.

The accurate description of the velocity and pressure fields upon impact
requires that the deformation of the droplet be captured as it falls. To allow
the droplet to equilibrate with the gas flow around it the droplet first falls
through a large simulation box of $1 \times 5 \si{\milli\m}$ at a relatively low
resolution of $320 \times 1600$ grid cells. Towards the wall, the mesh is
refined six times. Each refinement divides a grid cell into four, giving a
smallest grid size of $50 \si{\nano\m}$. As initial condition for $\alpha$, the
droplet is assumed to be a perfect sphere, with its center of mass located at a
height of $4.5 \si{\milli\m}$ above the surface. To reach faster convergence of
the gas velocity field, Stokes flow is assumed both inside and outside the
droplet as initial condition, and stresses are matched on the interface
\cite{batchelor2000}. The pressure field does not need to be initialized and
develops as the simulation progresses. Once the droplet is sufficiently close to the wall, the
simulation results in the lower part of the large simulation box
are saved and re-initialized within a small box of $1 \times 0.5 \si{\milli\m}$
with a resolution of $320 \times 160$ grid cells which again gets refined towards the
wall. The results presented in this work are from simulations at two different
refinement levels. To capture the details at the contact line and to capture
splashing it is sufficient to refine the mesh eight times at the wall, for a minimum grid size of about $12 \si{\nano\m}$. From a computational point of
view, this is the smallest feasible mesh size that we can adopt for this system
to run the simulations long enough to observe splashing. A second set
of simulations focuses only on very early times. Because they run
for a much shorter time it is possible to refine the mesh ten times and obtain an
even higher resolution of about $3 \si{\nano\m}$. Complete convergence at the
contact line would require a grid size below the slip length, which is approximately $1
\si{\nano\m}$ and is beyond the reach of our computational resources.
Nevertheless, at grid sizes of $3 \si{\nano\m}$ and $12 \si{\nano\m}$ the
necessary physics of splashing are already present, and we expect the main
observations of our simulations to be qualitatively correct. The width of both
boxes is chosen to be large enough for the splash to occur within their
confines.

For the simulations of a full splash the impact velocity is $v_{0} = 10.0
\si{\m.\s^{-1}}$ and the fluids are ethanol, for the liquid phase, and air, for the gas phase. This combination
has been used in experiments \cite{xu2005}, and has the advantage of showing an
early splash, i.e. one that occurs shortly after impact. In experiments, the pressure effect is
observed at moderately reduced gas pressures \cite{xu2005}. We therefore assume
the dynamic viscosity of the air to be constant \cite{kadoya1985}, but the density and the
kinematic viscosity are allowed to change as pressure is reduced
\cite{kadoya1985}. The reduced ambient pressure used in the simulations was
deliberately chosen to be significantly lower than the experimentally observed
threshold pressure, to ensure that the simulations were performed well into the
suppressed splashing regime.

The impact velocities for the simulations resolving the early time scales are
$v_{0} = 2.5 \si{\m.\s^{-1}}$, $v_{0} = 5.0 \si{\m.\s^{-1}}$, and $v_{0} = 10.0
\si{\m.\s^{-1}}$. The liquid viscosity is either the viscosity of ethanol or ten
times that of ethanol and the surface tension is either the normal surface
tension between air and ethanol or twenty times this surface tension, which is
of the same order as the surface tension of mercury in air. In order to keep
memory requirements within the constraints of our infrastructure, we consider a
two-dimensional axisymmetric droplet with a diameter of $300 \si{\micro\m}$ (as
opposed to the $3 \si{\mm}$ droplets used in experiments \cite{xu2005}).

 
\section{Results}
 

\subsection{Overview}

\begin{figure*}[htpb]
\centering
\includegraphics[width=\textwidth]{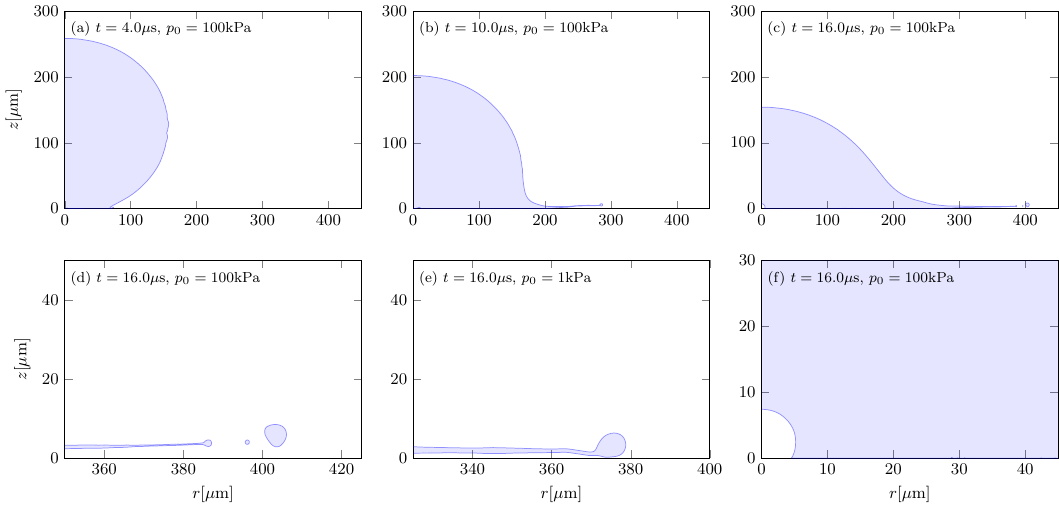}
\caption{Time series of the impact of a droplet on a dry surface. The top frames
(a-c) show simulation results for atmospheric pressure at different times. The
bottom frames show (d) liquid sheet breakup at atmospheric pressure, (e) no
breakup at reduced pressure, and (f) the central air bubble.}
\label{fig:timeSeriesSheetSplash} 
\end{figure*}
To provide a context for the next sections on the behavior of the gas film under
the droplet and lamella ejection, in this
section an overview is given of how the simulations reproduce the pressure
effect. Figure~\ref{fig:timeSeriesSheetSplash} shows a time series of the
pressure effect for a droplet approaching the surface at $v_{0} = 10
\si{m.s^{-1}}$. The top half shows three images of droplet impact and liquid
sheet ejection on a completely wetting surface at ambient pressure, $p_{0} = 100
\si{\kilo \pascal}$. The first frame, figure~\ref{fig:timeSeriesSheetSplash}a,
shows the droplet right after impact. As the droplet approaches the wall, gas
pressure builds up at the stagnation point to about $p = 1200 \si{\kilo
\pascal}$. This causes the droplet to deform, and a gas film in the shape of
a spherical dome to appear underneath it. As the gas film becomes thinner, air gets squeezed out and reaches velocities of up to
$\abs{\vec{v}} = 150 \si{\m.\s^{-1}}$. When the liquid eventually touches the
wall, a small amount of air is permanently trapped, forming the central air
bubble with a diameter of about $d = 10 \si{\micro \m}$
\cite{chandra1991,driscoll2011,mandre2012}. This bubble can be seen in
figure~\ref{fig:timeSeriesSheetSplash}f.

Right after impact a liquid sheet is ejected with a velocity of approximately
$v_{e} = 100 \si{\m.\s^{-1}}$. The liquid sheet forms at both atmospheric, $p_{}
= 100 \si{\kilo \pascal}$, and reduced pressure, $p_{0} = 1 \si{\kilo \pascal}$.
However, at atmospheric pressure the sheet gets lifted and breaks up into
smaller droplets; in contrast, at reduced pressure it stays close to the surface
and remains intact. Figure~\ref{fig:timeSeriesSheetSplash}c shows the liquid
sheet as it breaks up into smaller droplets at atmospheric pressure.
Figure~\ref{fig:timeSeriesSheetSplash}d shows a magnified image of the liquid
sheet as it breaks up. The droplet that breaks off has a diameter of $d = 7
\si{\micro \m}$. Figure~\ref{fig:timeSeriesSheetSplash}e shows a snapshot taken
at the same time, but at a reduced pressure. One can see that instead of
breaking up, the liquid sheet stays in one piece: splashing is suppressed by
decreasing the ambient pressure. 

\begin{figure}[htpb]
\centering
\subfloat
{\includegraphics[width=0.23\textwidth]{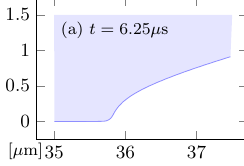}\label{fig:timeSeriesLamella}}
\subfloat
{\includegraphics[width=0.23\textwidth]{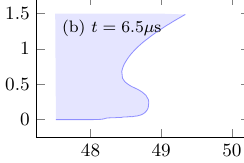}\label{fig:timeSeriesSheet}} \\
\subfloat
{\includegraphics[width=0.23\textwidth]{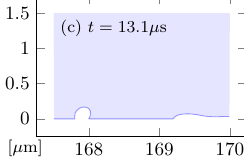}\label{fig:timeSeriesEntrainment}}
\subfloat
{\includegraphics[width=0.23\textwidth]{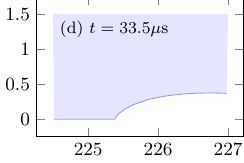}\label{fig:timeSeriesStable}} \\
\caption{Time series of the droplet interface showing the evolution of the
contact line at atmospheric pressure. Light blue represents the liquid phase.
The vertical axis shows the distance normal to the surface and the horizontal
axis shows the distance parallel to the surface relative to the center of the
droplet. Both axes are in $\si{\micro\m}$.} 
\label{fig:timeSeriesContact}
\end{figure}
The reason simulations are useful to investigate the gas film behavior and liquid
sheet ejection is that they provide access to
length and time scales that are not easily observed in experiments. This feature
is illustrated in figure~\ref{fig:timeSeriesContact}, which presents a
magnified time series of the liquid/gas interface at the contact line. A blue
color corresponds to the liquid phase and white to the gas phase. These
snapshots are taken at atmospheric pressure, from the moment just after impact
until the point when the edge of the droplet is about to leave the simulation
box. Drop impact is defined as the moment that a droplet would have hit the
surface continuing its trajectory instead of spreading on a gas film. This
occurs at $t=6.2 \si{\micro\s}$. Figure~\ref{fig:timeSeriesLamella} shows the
contact line just after impact and right before a liquid sheet forms at $t=6.25
\si{\micro\s}$. Once the liquid sheet forms, which can be seen in
figure~\ref{fig:timeSeriesSheet}, it is ejected right away. A feature revealed
by experiments \cite{driscoll2011} is contact line dewetting. At high speeds the
interface in front of the contact line becomes unstable and touches down on the
surface, causing gas bubble entrainment at the contact line. This phenomenon can
be appreciated in figure~\ref{fig:timeSeriesEntrainment}, which at $r \approx
168 \si{\micro m}$ shows a gas bubble that formed when the contact line became
unstable, and the liquid in front of the contact line touched down on the
surface. The first touch-down event is observed at $t=7.33 \si{\micro\s}$, and
at $t=14.4 \si{\micro\s}$ the contact line stabilizes again.
Figure~\ref{fig:timeSeriesEntrainment} and figure~\ref{fig:timeSeriesStable}
are well within the unstable and stable contact-line regimes, respectively.

Resolving the ejection of this same lamella in an experiment with a high-speed camera, would call for the
following specifications: considering a deacceleration of about $-4.0 \times
10^{7} \si{m.s^{-2}}$ right after ejection, to capture a velocity difference of
$8 \si{m.s^{-1}}$, 10\% of the ejection velocity, one would need to capture one
frame every $0.2 \si{\mu s}$. This corresponds to a frame rate of $5.000.000$
frames per second. To resolve a lamella thickness of $0.5 \si{\micro\m}$
with $10$ pixels, a resolution of $0.05 \si{\micro\m}$ per pixel is needed.
Methods like Total Internal Reflection (TIR) \cite{kolinski2014}, and
interference \cite{driscoll2010} can achieve very high spatial and temporal accuracy. However, only a
very small part of the droplet interface can be studied, while simulations show
the complete interface, including the gas film and the lamella.
 

\subsection{Gas film}
\label{seq:gasFilm}

\begin{figure*}[htpb]
\includegraphics[width=\textwidth]{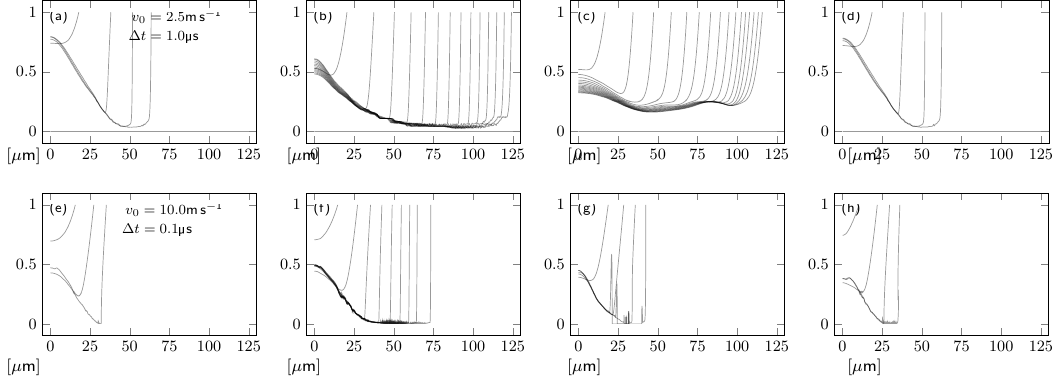}
\caption{Gas film profiles of different droplets approaching the wall. The
vertical axis shows the height above the wall and the horizontal axis the radial distance away
from the center. Both axes are in $\si{\micro \m}$. The figures in the top row
((a), (b), (c) and (d)) show an impact velocity of $2.5 \si{\m.\s^{-1}}$ and the
time difference between successive lines is $1.0 \si{\micro \m}$. The figures in
the bottom row ((e), (f), (g) and (h)) show an impact velocity of $10.0
\si{\m.\s^{-1}}$ with a time difference of $0.1 \si{\micro \m}$. Figures (a) and
(e) show the results for ethanol at atmospheric ambient pressure, figures (b)
and (f) show the results for the high viscosity liquid (i.e. a viscosity $10$ times higher
than the viscosity of ethanol), figures (c) and (g)
show the results for the liquid with a high surface tension (i.e. a surface
tension $20$ times higher than the surface tension of ethanol in air), and figures (d) and
(h) show the results for ethanol at reduced ambient pressure (i.e. an ambient
pressure $1/100$ of atmospheric pressure).}
\label{fEvolution}
\end{figure*}
This section looks in more detail at the gas film which forms under the droplet
upon impact. The evolution of the gas film for various parameters is shown in
figure~\ref{fEvolution}. The vertical axis shows the height away from the wall
and the horizontal axis the radial distance away from the center. Both axes are
in $\si{\micro \m}$. The figures in the top row show an impact velocity of $2.5
\si{\m.\s^{-1}}$ and the time difference between successive lines is $1.0
\si{\micro \m}$. The figures in the bottom row show the gas film for an impact
velocity of $10.0 \si{\m.\s^{-1}}$ with a time difference of $0.1 \si{\micro
\m}$ between successive lines. Figures (a) and (e) show the gas film for ethanol
at atmospheric ambient pressure, figures (b) and (f) show the results for the
high viscosity liquid (i.e. a viscosity $10$ times higher
than the viscosity of ethanol), figures (c) and (g) show the results for the liquid
with a high surface tension (i.e. a surface
tension $20$ times higher than the surface tension of ethanol in air), and figures (d) and (h) show the results for
ethanol at reduced ambient pressure (i.e. an ambient
pressure $1/100$ of atmospheric pressure). Comparing the top and bottom row
shows that a higher impact velocity results in a thinner gas film. 
The fluctuations of the interface close to the wall in the bottom row are the
result of the gas film collapsing behind the edge of the spreading droplet.
Comparing figure (a) and (b), shows that for early times a higher viscosity
results in a thicker gas film. Both the effect of velocity and viscosity are
consistent with literature~\cite{kolinski2014}. An increases surface tension
also gives an increased gas film thickness while a lower ambient pressure has
little effect on the thickness of the gas film.

\begin{figure}[htpb]
\includegraphics[width=0.48\textwidth]{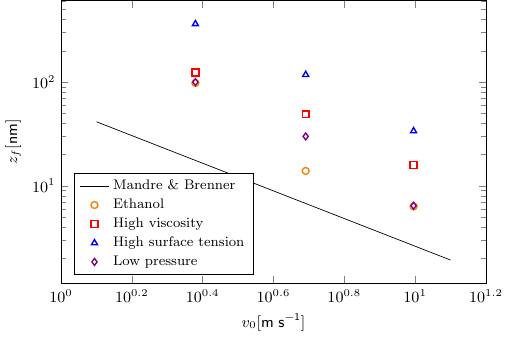}
\caption{Height of the gas film under the drop at the moment of lamella
formation for various velocities, viscosities, and surface tensions. The line
shows the theoretical predictions by \citet{mandre2012}.}
\label{vImpactvszFilm}
\end{figure}
A more quantitative representation of the gas film height can be seen in
figure~\ref{vImpactvszFilm}. This figure shows the height of the gas film at the
edge of the droplet in $\si{\nano\m}$ at the moment of lamella formation as a function of impact
velocity. The observations made above for an impact velocity of $2.5
\si{\m.\s^{-1}}$ concerning the effect of viscosity, surface tension, and
pressure on the thickness of the gas film are shown to also hold for higher impact
velocities. Moreover, an higher impact velocity can be seen to reduce the
thickness of the gas film irrespective of the liquid properties of the droplet
or the ambient gas pressure. In addition to the various data points,
figure~\ref{vImpactvszFilm} also shows a prediction by \citet{mandre2012} for
the thickness of the gas film as function of impact velocity. They propose a
scaling of $z_{f} = 60 r_{0}
\txt{St}^{4/3}$, where $\txt{St} = \mu_{g} / (\rho_{l} v_{0} r_{0})$ is the
Stokes number, $r_{0}$ is
the radius of the droplet, $\rho_{l}$ is the liquid density, and $\mu_{g}$ is
the gas viscosity. While the prefactor is different, possibly due to differences in geometry and the fact that the
theory was developed for inviscid flow, the scaling is consistent with theory.

\begin{figure}[htpb]
\includegraphics[width=0.48\textwidth]{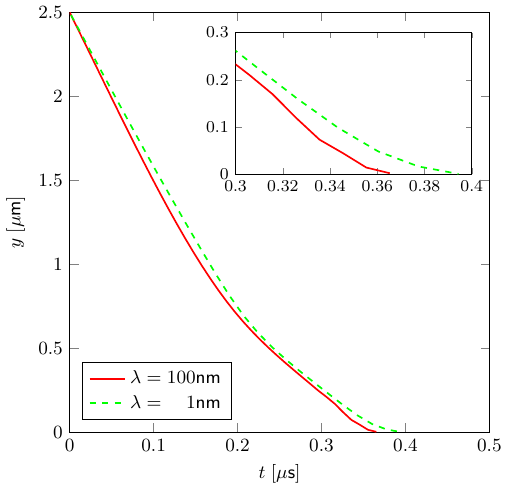}
\caption{The minimum thickness of the gas film as function of time comparing two
simulations: one simulation with a slip length of $\lambda = 1 \si{\nano\m}$, and
one simulation with a slip length of $\lambda = 100 \si{\nano m}$. Both
simulations are performed at reduced ambient pressure. The inset shows a
magnified view of the droplets approaching the wall.} 
\label{fig:slipLength}
\end{figure}

One aspect of the figure that needs to be examined more closely is that
fact that, especially at higher impact velocities, the thickness of the gas film
under the droplet upon impact becomes very thin. Considering that at room
temperature and atmospheric pressure the mean free path is around $\lambda
\approx 70 \si{\nano\m}$ \cite{jennings1988}, this suggests that the continuity
assumption is not valid and raises the question how a breakdown of the continuity assumption would affect the
simulation results. \citet{mandre2012} show that at larger impact velocities the
continuum assumption breaks down and they predict that this leads to earlier
rupture of the gas film. The effect of the breakdown of continuity can be
modeled by increasing the slip length on the wall \cite{sprittles2017}.
Figure~\ref{fig:slipLength} shows the minimum thickness of the
gas film as function of time comparing slip lengths of $1 \si{\nano m}$ and $100
\si{\nano m}$. Both simulations are of an ethanol droplet falling through air at
reduced ambient pressure. The simulation results are only show till right
before collapse of the gas film. The code supports only one slip length and as
mentioned in the theory \& methods section the slip length of $\lambda = 1
\si{\nano m}$ was chosen because it is a good value for the contact line. The
slip length of $\lambda = 100 \si{\nano m}$ was chosen because the slip length is
inversely proportional to pressure \cite{sprittles2017} and the ambient pressure
was reduced hundred fold. It can be appreciated that the increased slip length does result in
a slightly earlier collapse of the gas film, consistent with literature
\cite{mandre2012}. In addition, it has been reported that at atmospheric ambient
pressure, the maximum contact line velocity is not greatly affected by the
presence of a slip boundary condition for the gas phase \cite{sprittles2017}. This suggests that
although the gas film is expected to collapse sooner when the continuum
assumption breaks down, there should be a gas film present at the edge of the
spreading droplet. 

\begin{figure}[htpb]
\centering
\subfloat
{\includegraphics[width=0.23\textwidth]{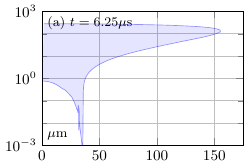}\label{fig:timeSeriesSheetSemiLog1}}
\subfloat
{\includegraphics[width=0.23\textwidth]{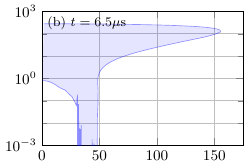}} \\
\subfloat
{\includegraphics[width=0.23\textwidth]{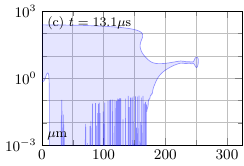}}
\subfloat
{\includegraphics[width=0.23\textwidth]{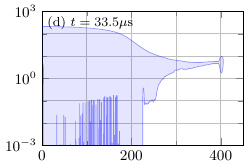}\label{fig:timeSeriesSheetSemiLog4}} \\
\caption{Time series of the droplet interface with the vertical axis plotted on
a logarithmic scale. The axis are in $\si{\micro\m}$ and the vertical axis is plotted on a
logarithmic scale. Since the vertical axis cannot go to zero, it is cut off at
$10^{-3} \si{\micro\m}$.}
\label{fig:timeSeriesSheetSemiLog}
\end{figure}

As mentioned in the introduction, apart from the gas film forming under the droplet upon impact there is also a
gas film under the liquid sheet. To be able to analyze the thickness of this gas
film, figure~\ref{fig:timeSeriesSheetSemiLog} shows a time series of the contour of
the droplet at atmospheric pressure with the vertical axis plotted on a
logarithmic scale. Since the vertical axis cannot go to zero, it is cut off at
$10^{-3} \si{\micro\m}$, and the center of the drop corresponds to the vertical
axis. Figure~\ref{fig:timeSeriesSheetSemiLog1} shows the droplet right after impact
with the central air dome clearly visible. Initially the bubble is slightly
under $1 \si{\micro\m}$ high, and about $30 \si{\micro\m}$ wide. The vertical
lines or peaks under the droplet are small air bubbles which were trapped on
impact. The next figure shows the droplet right after a liquid sheet is ejected
and the third frame shows the droplet in the unstable contact line regime. The
vertical lines behind the contact line are small air bubbles which were
entrained by the moving contact line. Also, it can be observed how the contact
line of the central air bubble is moving toward the center. The gas layer under
the liquid sheet is on the order of several microns thick. The last figure,
figure~\ref{fig:timeSeriesSheetSemiLog4}, shows the droplet with a stable
contact line. There are no longer droplets being trapped right behind the
contact line. Instead, there is just one cavity present at the contact line with
a height of a little under $1 \si{\micro\m}$. The gas film right after the
cavity has a thickness of about $0.1 \si{\micro\m}$, and increases again to
several microns towards the rim of the liquid sheet.

To put the above numbers in perspective they need to be compared to the mean
free path of air. At room temperature and atmospheric pressure the mean free
path is around $\lambda \approx 70 \si{\nano\m}$ \cite{jennings1988}. The typical thickness
of the gas layer is several microns, so the continuity approximation should be
valid for these simulations. Since the mean free path scales with the kinematic
viscosity, at the reduced pressure the mean free path is about $\lambda \approx 7
\si{\micro\m}$, which is about the same height as the typical thickness of the
gas layer. This means that at reduced pressure the validity of
the continuum approximation is pushed to its limits, and that the simulations
may slightly overestimate the longevity of the gas film under the liquid
sheet at reduced pressure. Additionally, at reduced ambient pressure the maximum
velocity of the contact line is expected to increase \cite{sprittles2017}. If
the liquid sheet is not ejected fast enough, this might affect the simulation
results.

\begin{figure}[htpb]
\centering%
\begin{minipage}[b]{0.5\textwidth}
\subfloat%
{\includegraphics[width=0.46\textwidth]{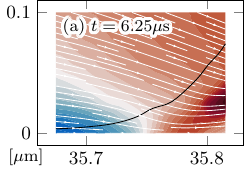}\label{fig:timeSeriesHighSpeed}}%
\subfloat%
{\includegraphics[width=0.46\textwidth]{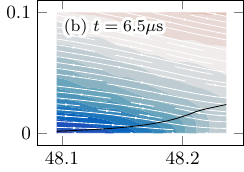}} \\%
\subfloat%
{\includegraphics[width=0.46\textwidth]{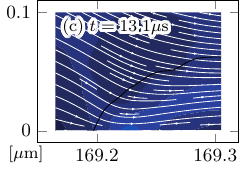}}%
\subfloat%
{\includegraphics[width=0.46\textwidth]{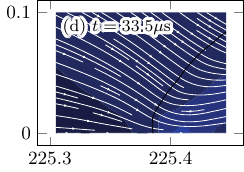}\label{fig:timeSeriesLowSpeed}} \\%
\end{minipage} \\
\begin{minipage}[b]{0.5\textwidth}
\hspace*{0.5cm}
\begin{turn}{-90}
\includegraphics[height=0.9\textwidth]{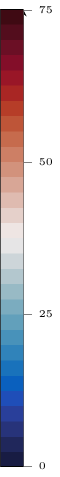}
\end{turn}
\end{minipage}
\caption{Time series of the droplet interface, velocity magnitude, and streamlines, showing the
evolution of the contact line at atmospheric pressure. This time series is
magnified $15$ times compared to figure~\ref{fig:timeSeriesContact}. The
velocity magnitude is in $\si{\m.\s^{-1}}$, and the axes are in \si{\mu\meter}.
The small arrows on the streamlines indicate the direction of the flow. 
}
\label{fig:timeSeriesSlip}
\end{figure}
\begin{figure}[htpb]
\centering
\begin{minipage}[b]{0.5\textwidth}
\subfloat
{\includegraphics[width=0.46\textwidth]{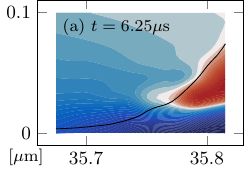}\label{fig:vorticitya}}
\subfloat
{\includegraphics[width=0.46\textwidth]{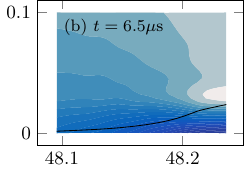}\label{fig:vorticityb}} \\
\subfloat
{\includegraphics[width=0.46\textwidth]{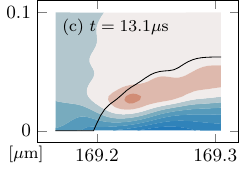}\label{fig:vorticityc}}
\subfloat
{\includegraphics[width=0.46\textwidth]{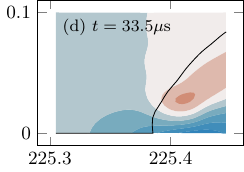}\label{fig:vorticityd}} \\
\end{minipage} \\
\begin{minipage}[b]{0.5\textwidth}
\hspace*{0.5cm}
\begin{turn}{-90}
\includegraphics[height=0.9\textwidth]{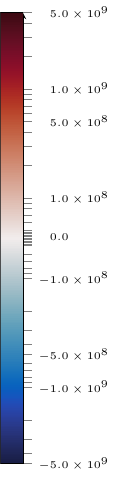}\label{fig:vorticityBar}
\end{turn}
\end{minipage}
\caption{Time series of the droplet interface and vorticity, $\omega$, showing
the evolution of the contact line at atmospheric pressure. This time series is
magnified $15$ times compared to figure~\ref{fig:timeSeriesContact}. The axes are
in \si{\mu\meter} and the contours in \si{\s^{-1}}.}
\label{fig:timeSeriesVorticity}
\end{figure}
Having addressed the issue of continuum assumption breakdown, an important
consequence of the existence of a thin gas film under the droplet can now be
discussed. Due to the the semi-logarithmic scaling of
figure~\ref{fig:timeSeriesSheetSemiLog} the contact line is difficult to see,
but in figure~\ref{fig:timeSeriesSlip} a magnified time series of the contact line and
corresponding streamlines is shown at atmospheric pressure.
Figure~\ref{fig:timeSeriesVorticity} shows the contact line and corresponding
vorticity contour plot, $\omega = \nabla \times \vec{v}$, at the same
magnification. While in the last two frames of both figures a normal low-speed
contact line can be observed, the first two frames show a strongly curved
interface, which continues under the droplet. To gain more insight into the
behavior of the curved interface at the contact line, we now focus on the
streamlines in figure~\ref{fig:timeSeriesSlip}. It can be appreciated that in
the first two frames, for a length of about $0.1 \si{\mu \m}$ along the wall, a
fluid parcel starting at the interface will closely approach the wall when tracing its
streamline. This behavior is reminiscent of the interface rolling over the
surface at a high speed, like a caterpillar track. The lower two frames show a
traditional sliding contact line and the area where a fluid parcel would end up
at the wall is much smaller. Looking at the vorticity contours in
figure~\ref{fig:timeSeriesVorticity}, a very strong negative vorticity can be
observed at the wall in the top two frames. This suggest that in this area fluid
parcels have a strong tendency to rotate clockwise as they move along their
streamlines, again suggesting the presence of a rolling motion. The presence of a rolling contact-line regime seems to be crucial to
splashing. Coarsening the resolution just enough so that the gas film at
the edge of the drop and the rolling contact line do not get resolved, prevents
a droplet from splashing on impact. It has been brought to our attention that
the rolling contact line regime recently also has been observed in another
publication \cite{philippi2016}.


\subsection{Lamella}

\begin{figure*}[htpb]
\includegraphics[width=\textwidth]{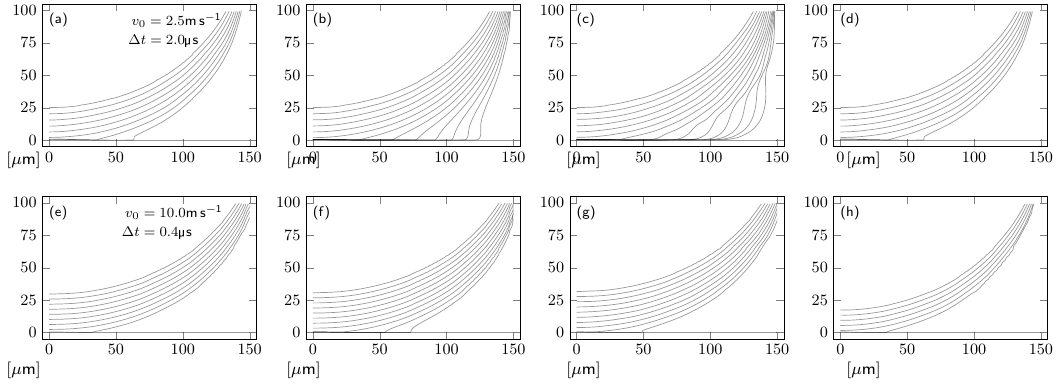}
\caption{Lamella profiles of different droplets approaching the wall. The
vertical axis shows the height above the wall and the horizontal axis the radial distance away
from the center. Both axes are in $\si{\micro \m}$. The figures in the top row
((a), (b), (c) and (d)) show an impact velocity of $2.5 \si{\m.\s^{-1}}$ and the
time difference between successive lines is $2.0 \si{\micro \m}$. The figures in
the bottom row ((e), (f), (g) and (h)) show an impact velocity of $10.0
\si{\m.\s^{-1}}$ with a time difference of $0.4 \si{\micro \m}$. Figures (a) and
(e) show the results for ethanol at atmospheric ambient pressure, figures (b)
and (f) show the results for the high viscosity liquid (i.e. $10$ times higher
than the viscosity of ethanol), figures (c) and (g)
show the results for the liquid with a high surface tension (i.e. $20$ times
higher than the surface tension of ethanol in air), and figures (d) and
(h) show the results for ethanol at reduced ambient pressure (i.e. $1/100$ 
of atmospheric ambient pressure).
\label{fig:lEvolution}}
\end{figure*}
Just like the previous section looked into the effect of various material
properties on the formation of a gas film under the droplet, this section looks
at the effect of these same parameters on lamella formation.
Figure~\ref{fig:lEvolution} shows various time series of droplets with impact
velocities of $2.5 \si{\m.\s^{-1}}$ (top row) and $10.0 \si{\m.\s^{-1}}$ (bottom
row). Figures (a) and (e) show ethanol droplets at atmospheric ambient pressure,
figures (b) and (f) show the droplets with a high viscosity of $10$ times the
viscosity of ethanol, figures (c) and (g) show the droplets with a high surface
tension of a fluid with $20$ times the surface tension of ethanol in air, and figures (d) and (h) show the
evolution of ethanol droplets at an ambient pressure reduced $100$ times
compared to atmospheric pressure. The droplets are
shown till the moment of lamella formation. It can be seen that an higher impact
velocity results in a thinner lamella which forms earlier. In addition, it can
be observed that both increased viscosity and surface tension results in thicker
lamella which are ejected later. Pressure on the other hand does not have an effect on the moment of lamella ejection.

\begin{figure}[htpb]
\includegraphics[width=0.48\textwidth]{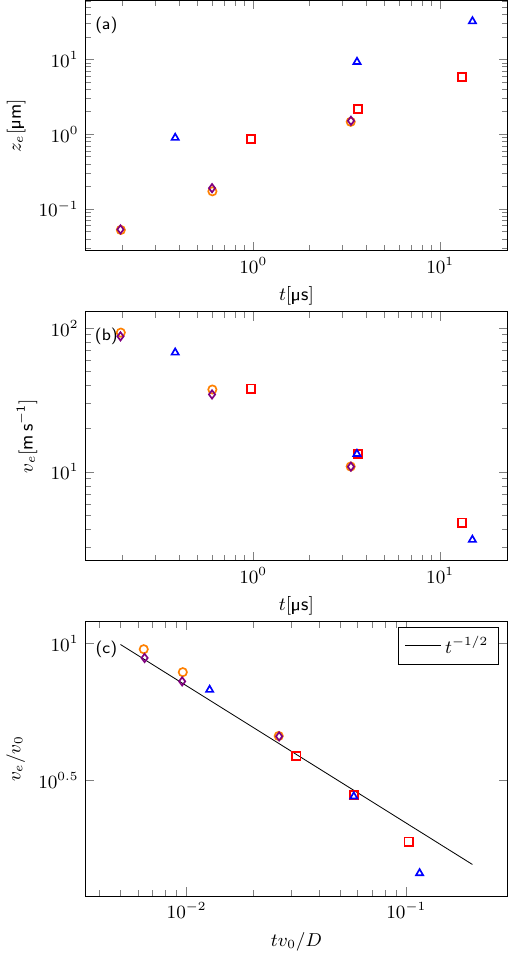}
\caption{Various lamella properties as a function of their ejection time. In all
three plots the results are shown for an ethanol droplet in a gas at atmospheric ambient pressure (
\protect\includegraphics{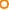}
), and ethanol drop in a gas at reduced ambient pressure (
\protect\includegraphics{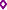}
), a droplet with a high surface tension in a gas at atmospheric ambient pressure (
\protect\includegraphics{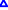}
), and a droplet with a high viscosity (
\protect\includegraphics{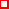}
). (a) Lamella height, $z_{e}$, as a function of ejection time, $t_{e}$. The
three data points per symbol correspond to impact velocities of $2.5
\si{\m.\s^{-1}}$, $5.0 \si{\m.\s^{-1}}$, and $10.0 \si{\m.\s^{-1}}$ with $10.0
\si{\m.\s^{-1}}$ resulting in the smallest ejection time. (b) Lamella ejection
velocity, $v_{e}$, as function of ejection time, $t_{e}$. The impact velocity can be
non-dimensionalized using the impact velocity, $v_{0}$, and droplet
diameter, $D$. The result of which can be seen in figure (c). The data follows a
scaling of $v_{e}/v_{0} \propto t^{-1/2}$.}
\label{fig:timeVsLamella}
\end{figure}
In figure~\ref{fig:timeVsLamella} the simulation data is shown in a more
quantitative form. Figure~\ref{fig:timeVsLamella} (a) shows the height of the
lamella, $z_{e}$, at the moment of ejection as a function of the time since
impact, $t$, on a log-log plot. The data is shown for a droplet of ethanol in both atmospheric and
reduced ambient pressure, a high viscosity droplet, and a high surface tension
droplet. The impact velocities are $2.5 \si{\m.\s^{-1}}$, $5.0 \si{\m.\s^{-1}}$,
and $10.0 \si{\m.\s^{-1}}$. A higher impact velocity results in an earlier
ejection time. An earlier ejection time, in turn, results in a thinner lamella.
Increasing the surface tension and viscosity delays the moment of lamella
ejection, resulting in a thicker lamella. In
figure~\ref{fig:timeVsLamella} (b) the lamella ejection velocity as function of
ejection time can be seen. For all the different liquid properties a later
ejection time gives a lower ejection velocity. In addition, when the data is
non-dimensionalized with the impact velocity, $v_{0}$, and droplet diameter, $D$,
the data collapses to a single line with an exponent of $v_{e}/v_{0} \propto
t^{-1/2}$. For especially the last point a deviation from this scaling can be
observed. The reason for this can be seen in figure~\ref{fig:lEvolution} (c). At the impact velocity
of $2.5 \si{\m.\s^{-1}}$ the increased surface tension slows down ejection time
by so much that the lamella size can not be considered much smaller than the
droplet size anymore, $z_{e} \not\!\ll D$, which results in different dynamics
for lamella formation.

\begin{figure}[htpb]
\includegraphics[width=0.48\textwidth]{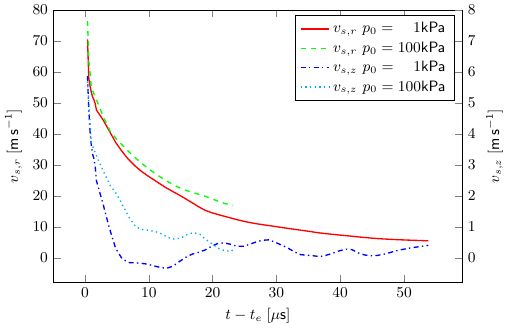}
\caption{The radial and vertical velocity of the liquid sheets ejected from the
droplet as function of time. $t_{e}$ is defined as the moment of sheet ejection.
\label{fig:velocities}}
\end{figure}
Simulations can also also provide insight on how the velocity of the lamella
evolves after ejection as it gets lifted up and becomes a liquid sheet.
Two simulations were continued after lamella formation. In both simulations the
droplet has an impact velocity of $v_{0} = 10 \si{\m.\s^{-1}}$, but one
simulation is at atmospheric ambient gas pressure and one is at reduced ambient
gas pressure. The results are shown in figure~\ref{fig:velocities}. The left
axis shows the radial liquid sheet velocity while the right axis shows the
vertical liquid sheet velocity. The velocity curves for the simulation at
atmospheric ambient pressure end when the liquid sheet breaks up. In this graph,
$t = 0$ is the moment of liquid sheet ejection, $t_{e}$, which is defined as the
moment a local maximum can first be detected in the width of the droplet close
to the wall. This local maximum is the beginning of a lamella. It can be
appreciated that both the initial vertical and horizontal ejection velocities
are very similar between the normal and reduced pressure cases. This is expected
because lamella ejection occurs very shortly after impact, and impact dynamics
are typically described by the Reynolds number and the Weber number
\cite{josserand2016}, neither of which depend on the properties of the ambient
gas. Only at later times do the velocities start to differ resulting in the
liquid sheet at atmospheric ambient pressure being lifted higher than the liquid
sheet at reduced ambient pressure, resulting in its breakup.
 
 
\section{Discussion}

While in this section various splashing and droplet deposition models are
discussed, the reproduction of the pressure effect in our simulations in itself leads to
important conclusions. To reduce the number of variables, both the liquid
and gas phases are assumed to be incompressible; this indicates that a shock
wave cannot be solely responsible for liquid sheet formation \cite{driscoll2011}. Also,
vortex formation in the gas phase upon impact is not necessary for splashing
\cite{bischofberger2013}. Although the vortices are not resolved in our
simulations, the effect of pressure is still captured. Lastly, the break up of
the liquid sheet is thought to be caused by either a Plateau-Rayleigh
instability, or a Rayleigh-Taylor instability \cite{agbaglah2013}. Although a
Plateau-Rayleigh instability can play a role in three-dimensional splashing,
since in these two-dimensional simulations breakup is still observed, it can be
concluded that such an instability is not a necessary condition to observe splashing.

\begin{figure}[htpb]
\includegraphics[width=0.48\textwidth]{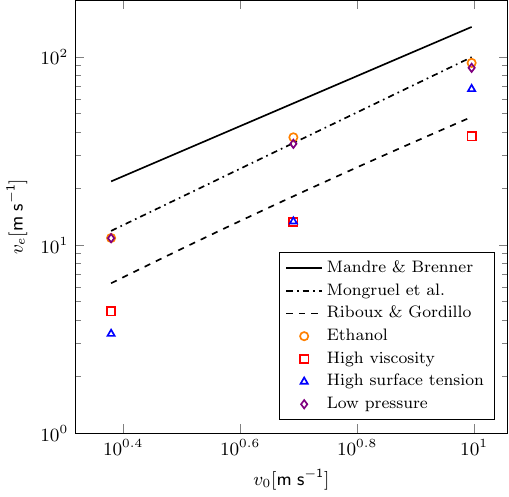}
\caption{Lamella ejection velocity as function of the impact velocity of a
droplet for various velocities, viscosities, and
surface tensions. Ethanol refers to the simulations of ethanol in air, ``High
viscosity'' refers to simulations of a fluid with $10$ times the viscosity of
ethanol in air, and ``High surface tension'' refers to simulations of a fluid with
$20$ times the surface tension of ethanol in air. The lines show theoretical
predictions by \citet{mandre2012} ($v_{e} \propto v_{0}^{4/3}$),
\citet{riboux2014} ($v_{e} \propto v_{0}^{1.3}$ in current regime), and
\citet{mongruel2009} ($v_{e} \propto v_{0}^{3/2}$). The error bars in this figure
are smaller than the symbols. \label{fig:vImpactvsvEject}}
\end{figure}

To illustrate the need to perform a detailed comparison between the various
existing models for splashing, in figure~\ref{fig:vImpactvsvEject} the ejection
velocity, $v_{e}$, is plotted as function of impact velocity, $v_{0}$. The
correct prediction of the ejection velocity is important, because it an
essential part of predicting splashing behavior \cite{riboux2014}. The
impact velocity is chosen here as the independent variable instead of time
because different models have different definitions for the time of impact, which becomes
relevant when ejection time occurs very shortly after impact. As was seen in
figure~\ref{fig:timeVsLamella}, it can be observed
that reducing the ambient pressure does not have much effect on lamella
formation. However, both an increased viscosity and an increased surface tension
significantly reduce the ejection velocity. In addition to the simulation data,
the graph shows theoretical curves corresponding to predictions for ethanol
drops in air. The formulas are shown in the different sections below. To plot the theoretical curves of \citet{mandre2012} and
\citet{riboux2014}, the prefactors that are provided with their respective
theories are used. Since, \citet{mongruel2009} do not provide a prefactor, a
prefactor of $0.275$ was chosen to match the data. To be able to clearly observe
the differences in scaling of the liquid sheet ejection velocity as function of
the impact velocity, velocities are not non-dimensionalized. 

It can be seen in this graph that different theories all predict a similar
scaling relation for the lamella ejection velocity as function of the impact
velocity. This makes it difficult to distinguish between these different
theories in both experiments and simulations, and in the following sections a
more fundamental look is taken at the assumptions that underlie each theory.

To begin with, the relevant non-dimensional numbers of the system are given
here:
\begin{align*}
\txt{St} &= \frac{\mu_{g}}{\rho_{l} v_{0} r_{0}}, & 
\txt{We} &= \frac{\rho_{l} v_{0}^{2} r_{0}}{\sigma}, \\
\txt{Re} &= \frac{\rho_{l} v_{0} r_{0}}{\mu_{l}}, &
\txt{and} \quad\quad\quad 
\txt{Oh} &= \frac{\sqrt{\txt{We}}}{\txt{Re}},
\end{align*}
where $\txt{St}$ is the inverse Stokes number, $\txt{Re}$ is the Reynolds
number, $\txt{We}$ is the Weber number, and $\txt{Oh}$ is the Ohnesorge number.
These numbers have the values: $\txt{St} = 1.3 \times 10^{-5}$, $\txt{We} =
528$, $\txt{Re} = 986$, and $\txt{Oh} = 2.33 \times 10^{-2}$, which where
calculated using the following parameters: $v_{0} = 10 \si{\m.\s^{-1}}$,
$\rho_{l} = 789 \si{\kg.\m^{-3}}$, $\mu_{l} = 1.20 \times 10^{-3}
\si{\pascal.\s}$, $\rho_{g} = 0.01 \si{\kg.\m^{-3}}$ at $p_{0} = 1 \si{\kilo
\pascal}$, $\rho_{g} = 1.00 \si{\kg.\m^{-3}}$ at $p_{0} = 100 \si{\kilo
\pascal}$, $\mu_{g} = 1.48 \times 10^{-5} \si{\pascal.\s}$, $\sigma = 0.02239
\si{\J.\m^{-2}}$, and $r_{0} = 150\si{\micro\m}$. These numbers suggest that in
our simulations impact is dominated by the inertia of the fluid.

\subsection{Dewetting}
In this section the dewetting model by \citet{riboux2014} is further
investigated. A key assumptions of this model is that the liquid moves fast
enough that the formation of a gas film under a liquid sheet is caused by the
liquid dewetting the surface. However, this is not consistent with the results
of the simulations. As shown in the results, instead of dewetting a rolling
contact line is observed. While the gas film under the droplet can collapse, the
edge of the droplet never touches the surface. Consistent with literature
\cite{liu2015}, the existence of the gas film seems essential for splashing.
Liquid sheet formation is suppressed when the gas film is not resolved in our
simulations. In addition, while this needs to be investigated further, the simulations with
different slip lengths hint at the possibility that at reduced ambient pressure
the slip length can become so large that the formation of the gas film is
suppressed \cite{sprittles2017}. This is also observed in experiments
\cite{driscoll2011}.

Since droplet deposition is dominated by inertia and not by contact line
dissipation \citep{squires2005}, independent of the argument of how a gas film
forms under the liquid sheet, one can investigate the model's predictions for
lamella formation and liquid sheet breakup. The theory states that a
lamella will be ejected when the acceleration of a material point in the lamella is
larger than the acceleration of the spreading radius of the drop. This
acceleration is determined by a balance between inertial, viscous, and
capillary forces, leading to an ejection time prediction:
\begin{equation}
  c_{1} \txt{Re}^{-1} t_{e,R}^{-1/2}
+ \txt{Re}^{-2} \txt{Oh}^{-2}
=
  c^{2} t_{e,R}^{3/2}
\label{eqn:t_eR}  
\end{equation}
Here $t_{e,R}$ is the sheet ejection time, non-dimensionalized with the impact
velocity, $v_{0}$, and the droplet radius, $r_{0}$, and $c_{1}$ and $c$ are two
constants with the values $c_{1} = \sqrt{3}/2$ and $c = 1.1$. Using the ejection
time, the ejection velocity can be calculated with the equation:
\begin{equation}
  v_{e,R}
=
  \frac{\sqrt{3}}{2} t_{e,R}^{-1/2}.
\label{eqn:v_eR}  
\end{equation}
The above equation is used to calculate the lamella ejection velocity in
figure~\ref{fig:vImpactvsvEject}. As was mentioned above, although the scaling
of the ejection velocity as function of the impact velocity is consistent with
our data, the provided fitting parameters do not fit our data well. This could
be explained by the fact that the experiments that were used to fit their model
did not have access to the time and length scales that these simulations can
capture. Since the velocity of a lamella changes rapidly after ejection, this
can lead to a significant error. However, even after fitting the model to our
simulation data for an ethanol droplet, we find that it correctly predicts the
effect of surface tension, but not the effect of viscosity.

After lamella formation the dewetting model predicts that a drop will splash or
not based on whether a lift force is able to lift up the liquid sheet into the
air, causing the drop to splash, or whether it re-wets the surface. This results
in the definition of a parameter $\beta = {v_{h_{t},z}}/{v_{\txt{tc}}}$, where
$v_{h_{t},z}$ is the vertical velocity of the liquid sheet when it has risen to
a characteristic height of $h_{t}$, and $v_{\txt{tc}}$ is the Taylor-Culick
velocity. In this work two different methods to calculate $\beta$ are used;
based on material properties and impact conditions, \citet{riboux2014} provide a
theoretical prediction for the value of $v_{h_{t},z}$. When this theoretical
value is used, $\beta$ is called $\beta_{R}$ and it is calculated using the
equation:
\begin{equation}
  \beta_{R}
=
  \brc{\frac{
    K_{l} \mu_{g} (v_{e,R} v_{0})
  + K_{u} \rho_{g} (v_{e,R} v_{0})^{2} (h_{t} r_{0})
  }{2 \sigma}}^{1/2}  
\end{equation}
with:
\begin{equation}
  K_{l} 
=
- \frac{6}{\tan^{2} {(\alpha)}} 
  \sqrbrc{\ln{(16 \frac{l_{g}}{h_{t} r_{0}})} - \ln{(1 + 16 \frac{l_{g}}{h_{t} r_{0}}})},
\end{equation}
and:
\begin{equation}
  K_{u} 
=
  0.3.
\end{equation}
In the above equation $\alpha = 20/180 \; \ang{\pi}$, and $l_{g} = 1.2 \lambda$, the slip
length of the gas. $\lambda = \lambda_{0} (p_{\txt{atm}}/p_{0})$ with $\lambda_{0}
= 65 \si{\nano\m}$ the mean free path of air at room temperature and atmospheric
pressure, $p_{\txt{atm}} = 100 \si{\kilo\pascal}$. However, $v_{h_{t},z}$ can also be
directly determined from the simulation data, in which case $\beta$ is called
$\beta_{S}$ and is calculated from our simulations using:
\begin{equation}
  \beta_{S}
=
  \frac{v_{h_{t},z} v_{0}}{\sqrt{2 \sigma / (\rho_{l} h_{t} r_{0})}}
\label{eqn:betaS}  
\end{equation}
where $v_{h_{t},z}$ is the vertical velocity of the liquid sheet when it has
risen a characteristic height of:
\begin{equation}
  h_{t} 
=
  \frac{\sqrt{12}}{\pi} t_{e,R}^{3/2}
\end{equation}
For $p_{0} = 1 \si{\kilo \pascal}$ this is the vertical velocity at $t=0.17$,
and for $p_{0} = 100 \si{\kilo \pascal}$ at $t=0.26$. 
\begin{table}[ht]
\centering
\caption{Comparison of simulation results with predictions of the model by
Riboux and Gordillo \cite{riboux2014} for two different ambient pressures. The table shows the ejection
velocities, and the splashing criterion of the liquid sheet. $v_{e}$, and
$\beta_{S}$ are simulation results. $v_{e,R}$, and $\beta_{R}$ are the
predictions by Riboux and Gordillo \label{riboux}}
\begin{tabular}{lrrrr}
\hline
\hline
$\mathbf{P_{0}} [\si{\kilo \pascal}]$ & 
$\mathbf{v_{e,R}}$   & 
$\mathbf{v_{e}}$     & 
$\mathbf{\beta_{R}}$ & 
$\mathbf{\beta_{S}}$ \\
\hline
$\phantom{00}1$ & $4.87$ & $7.05$ & $0.075$ & $0.523$ \\
$100$           & $4.87$ & $7.65$ & $0.649$ & $0.557$ \\
\hline
\hline
\end{tabular}
\end{table}
In Table \ref{riboux} the theoretical predictions of Riboux and Gordillo \cite{riboux2014} are shown. The theoretically predicted values for $\beta$ are
$\beta_{R} = 0.075$ and $\beta_{R} = 0.649$ for calculations at $p_{0} = 1
\si{\kilo \pascal}$ and $p_{0} = 100 \si{\kilo \pascal}$, respectively. However,
the values calculated from the simulations are $\beta_{S} = 0.523$ and $\beta_{S} = 0.557$. 
Comparing $\beta_{S}$ and $\beta_{R}$, shows that
theory predicts about an order of magnitude difference between the normal and reduced pressure
cases. However, the values calculated from the simulations show similar values
of $\beta$ for both normal and reduced pressures. This suggests that when using
the definition in equation~\ref{eqn:betaS}, $\beta$ is not able to predict
splashing for these simulations. However, the vertical velocity components,
shown in figure~\ref{fig:velocities}, do suggest that a lift force could be
responsible for the breakup of a liquid sheet. Consistent with
equations~\ref{eqn:t_eR} and \ref{eqn:v_eR} the lamella ejection time and
velocity are independent of pressure. Only at later times, when a lift force
could have started to act on the liquid sheet, do the velocities start to
differ, supporting the lift force hypothesis. In addition, the existence of a
lift force would explain why, although there is a strong vertical component to
the liquid sheet velocity in figure~\ref{fig:velocities}, the liquid sheet does
not rise as high as is typically seen in experiments. The droplet size used in
these simulations is much smaller than a typical droplet used in experiments
resulting in a smaller liquid sheet and thus smaller lift force. 

While the dewetting theory is able to correctly predict splashing for a large
number of experiments \cite{riboux2014,goede2017}, the theory's predictions do
not match well with the simulations presented in this work. This includes the
dewetting assumption for the formation of the gas film, the model for lamella
formation, and the splashing criterion. A possible explanation for the mismatch
between the simulations and theory is that the experiments that were used to
benchmark the fitting parameters in the theory could not have measured liquid
sheet velocities at the early times observed in our simulations. Another mayor
difference with comparisons where the dewetting theory gives good predictions
for lamella formation and splashing is that the typical droplet size in these
experiments is on the order of $\si{\milli\m}$ while the droplets in this work are significantly
smaller. \citet{visser2015} also find that the dewetting model does not
correctly predict splashing for their micro-droplets and attribute this
to violating the assumption that the air film thickness under the liquid
sheet exceeds the mean free path length of the gas molecules. However, as
shown above, in this work at atmospheric pressure the gas film under the liquid
sheet is thicker than the mean free path of the gas molecules, which leaves open
the question whether there are other size effects at play.

\subsection{Skating}
This section focuses on the stating droplet theory of \citet{mandre2012}. As
shown in section~\ref{seq:gasFilm}, the simulation data are consistent with the
basic assumption of the skating model that the droplet moves on top of a very
thin gas film upon impact. While this gas film can collapse with time, when a
liquid sheet forms the edge of the droplet never touches the surface. The thin
gas film has also been observed in recent experiments, both directly
\cite{kolinski2012} and indirectly \cite{liu2015}. In addition, the simulation
results are consistent with the proposed scaling of the non-dimensional gas film
height:
\begin{equation}
  z_{f} 
= 
  60 \txt{St}^{4/3}
\end{equation}
where $\txt{St} = \mu_{g} / (\rho_{l} v_{0} r_{0})$ is the Stokes number,
$r_{0}$ is the radius of the droplet, $\rho_{l}$ is the liquid density, and
$\mu_{g}$ is the gas viscosity. However, because the theory was developed for
inviscid flow, it does not account for the effect of viscosity and surface
tension. Also consistent with the theory the gas film
scaling does not depend on gas pressure. However, when in the simulations the slip length
on the wall is increased the gas film drains faster. Since the slip length
depends on the ambient gas pressure \cite{sprittles2017}, it would be
interesting to perform a follow-up study with a wider range of slip lengths to
study the effect of the breakdown of continuity on the gas film at low gas pressure.

The presence of the gas film and the introduction of a rolling contact line also
provides a possible explanation for the recent experimental finding that
splashing is independent of the wetting properties of the surface
\cite{latka2018}. When the contact line is in the rolling regime its contact
angle is always $180$ degrees, and thus it should behave in exactly the same
manner on wetting and non-wetting surfaces. Therefore, in this regime,
liquid-sheet formation is predicted to be independent of the wetting properties
of the surface.

The non-dimensional lamella ejection velocity can be calculated with the
following equation: 
\begin{equation}
v_{e,M} = 0.34 \txt{St}^{-1/3}
\end{equation}
where the liquid sheet ejection velocity, $v_{e,M}$, is non-dimensionalized with
the impact velocity, $v_{0}$. As was shown in figure~\ref{fig:vImpactvsvEject},
the predicted scaling of the ejection velocity as function of the impact velocity is
consistent with the simulation results. However, the prefactor does not match,
possibly due to geometry differences. Also, just as was the case for the
scaling of the gas film height, the effects of viscosity and surface tension are
not captured in this model.

In addition to scaling relations for the gas film and the lamella ejection
velocity, \citet{mandre2012} propose the hypothesis that splashing is caused by
liquid deflecting on the surface. If this were the case, one would expect
different ejection velocity directions for a droplet impacting at atmospheric
ambient pressure and at reduced ambient pressure. It can be observed in
figure~\ref{fig:velocities} that this is not consistent with the simulation data,
and that thus our results suggest that deflection on the surface is not valid
splashing mechanism.

\subsection{Lamella formation}

In this last section the model for lamella formation developed by
\citet{mongruel2009} is discussed. As was shown in
figure~\ref{fig:vImpactvsvEject}, the proposed scaling for the ejection
velocity:
\begin{equation}
  v_{e,N}
\propto
  \sqrt{\txt{Re}}
\end{equation}
matches the simulation data well. In addition, this model predicts correctly
that the ejection velocities for high viscosity droplets are about $\sqrt{10}$
times lower than the ejection velocities for the normal ethanol droplets.
However, this model does not include surface tension and thus does not
capture its effect.

\begin{figure}[htpb]
\includegraphics[width=0.48\textwidth]{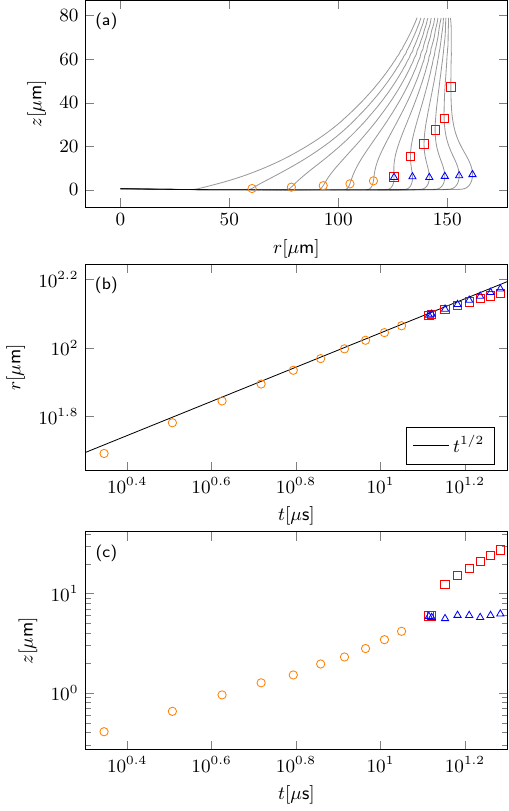}
\caption{a) Interface evolution of droplet impact at $2.5 \si{\m.\s^{-1}}$
with a liquid viscosity 10 times that of ethanol. A maximum in the horizontal
velocity
(
\protect\includegraphics{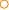}
) can be observed on the interface early after impact and can be traced
to the time of lamella formation. At the moment of lamella formation there is a
bifurcation for points following the cusp (
\protect\includegraphics{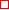}
) and the lamella (
\protect\includegraphics{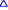}
). b) The radial position plotted as function of time for the same data points as in
figure~(a). The $t^{1/2}$ scaling suggests that the flow is dominated by inertia. c) The
vertical position of the velocity maximum, lamella and cusp as function of time.
\label{uEvolution}}
\end{figure}
One of the predictions of the model is that at early times viscosity is dominant
over inertia in the development of the lamella. To further explore the
effects of both inertia and viscosity, but also surface tension, in figure~\ref{uEvolution} we
examine the early stages of lamella formation. Figure~\ref{uEvolution}a shows
the interface of a droplet depositing on the surface with a velocity of $v_{0} =
2.5 \si{\m.\s^{-1}}$ and a viscosity of ten times that of ethanol. Between every
snapshot of the interface is a time difference of $\Delta t = 2 \si{\mu\s}$.
Right after impact, a radial velocity maximum (yellow dots) can be observed on
the interface; when we track the time evolution of this maximum, eventually a
lamella forms at the location of the velocity maximum. After the lamella forms,
when the radial position of the interface is plotted ($r$, as function of the
height, $z$), a neck or cusp appears that reflects a minimum of the interface
(red dots), and the lamella, which represents a maximum of the interface (blue
dots).

Figure~\ref{uEvolution}b shows the radial position of these points as a function
of time. The scaling with $r \propto t^{1/2}$ is a theoretical prediction based
on geometrical arguments and indicates that the radial positions of the velocity
maximum and lamella are dominated by inertia. This is consistent with the literature
\cite{mongruel2009}. Figure~\ref{uEvolution}c shows the
vertical positions of the data points as a function of time.
 
\begin{figure}[htpb]
\includegraphics[width=0.48\textwidth]{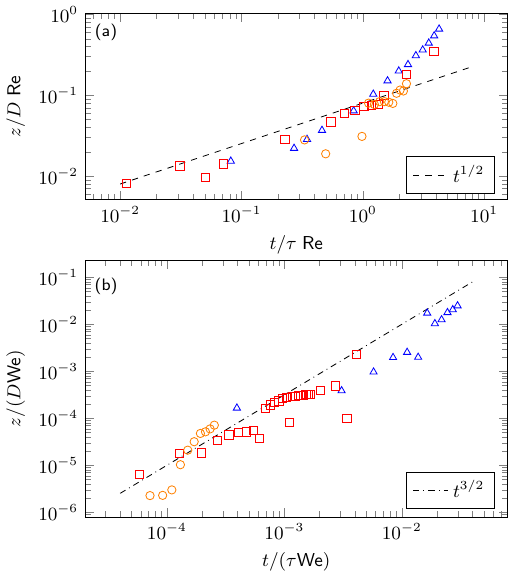}
\caption{a) Non-dimensional vertical position of the velocity maximum as function of
non-dimensional time for $2.5 \si{\m.\s^{-1}}$ 
(
\protect\includegraphics{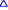}
), $5.0 \si{\m.\s^{-1}}$ (
\protect\includegraphics{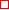}
), and $10.0 \si{\m.\s^{-1}}$ (
\protect\includegraphics{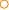}
). The surface tension is that of ethanol and the viscosity is
$10$ times that of ethanol. The scaling suggests that at early times, lamella formation is dominated by viscosity.
b) Non-dimensional vertical position of the velocity maximum as function of
non-dimensional time for $2.5 \si{\m.\s^{-1}}$ (
\protect\includegraphics{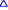}
), $5.0 \si{\m.\s^{-1}}$ (
\protect\includegraphics{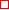}
), and $10.0 \si{\m.\s^{-1}}$ (
\protect\includegraphics{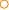}
). The viscosity is the viscosity of ethanol and the surface tension is $20$
times that of ethanol. The scaling suggests that at late times,
the scaling for these droplets is dominated by surface tension.
\label{scaling}}
\end{figure}

Figure~\ref{scaling} further explores the evolution of the velocity maximum
(i.e. the yellow dots in figure~\ref{uEvolution}) for various simulations. In
figure~\ref{scaling}a the results are shown for
droplets with a viscosity ten times that of ethanol at three
different impact velocities. Non-dimensionalizing both the vertical
position and time with the kinematic viscosity, $\nu$, and impact velocity, $v_{0}$,
leads to a collapse of the data at early times. This confirms that at early times lamella formation is dominated by
viscosity, as is predicted by \citet{mongruel2009}. Figure~\ref{scaling}b shows
the results of a droplet with a surface tension of twenty times that of ethanol impacting at three different velocities. This time the
axis is scaled using surface tension, $\sigma$, liquid density, $\rho$, impact
velocity, $v_{0}$, and diameter $D$. Again this leads to collapse of the data, suggesting that, for these droplets, late-stage
lamella formation is dominated by surface tension. The height of the lamella
scales as: $z_{\sigma} \propto \sqrt{\sigma v_{0} /(\rho D^{2})} t^{3/2}$.
The fact that the same scaling of the lamella height before ejection as function
of time can be observed for different material properties
in figure~\ref{fig:vImpactvsvEject}, and for different velocities in figure~\ref{uEvolution}
and figure~\ref{scaling}, suggests that a universal mechanism exists that causes
the lamella to be created for splashing and depositing droplets.
Capturing these different regimes could become a prerequisite for emerging splashing
theories, as opposed to a focus on
low-viscosity splashes \cite{mandre2012,riboux2014}.
 
 
\section{Conclusions}

In summary, high-resolution numerical simulations of splashing ethanol
droplets appear to describe the so-called ``pressure effect''
with considerable fidelity. The formation of the experimentally observed central
air bubble, air bubble entrainment at the contact line, liquid sheet formation,
and the scaling of the height of the gas film under the droplet with impact
velocity are all reproduced. 

The results of the simulations are compared with the dewetting theory of
\citet{riboux2014}, the skating droplet theory of \citet{mandre2012}, and the
lamella formation theory of \citet{mongruel2009}. Analyzing the gas film present
under the droplet upon impact no dewetting is observed. Instead, the simulations
confirm that there is a thin gas film present under the
droplet upon impact. While this gas film is unstable and may collapse, is is always present at
the edge of the spreading droplet. This is caused by the contact line moving in
a rolling fashion, continuously extending the gas film behind it. The presence
of the gas film seems critical for splashing since under-resolving the gas film
results in the suppression of splashing. In addition, the results of the
simulation are consistent with the scaling of the gas film height as function of
impact velocity proposed by \citet{mandre2012}. However, the model does not
incorporate the effect of liquid viscosity or surface tension. An area for
future research is to study the non-continuum effects in the gas film by
investigating the effect of various slip lengths on drainage and collapse of the
gas film \cite{sprittles2017}.

While the lamella formation theory does not incorporate surface tension, it does
predict the correct scaling of the lamella ejection velocity as function of the
impact velocity for both ethanol and a high viscosity liquid. Further analysis
of the start of the lamella formation right after impact confirms the existence of
an early-time viscosity-dominated regime. Also, a new length scale can be
defined for the height of the lamella before ejection to incorporate the effect
of surface tension in the lamella formation model.

Concerning the breakup of the liquid sheet after ejection our results are
not consistent with the hypothesis that splashing is caused by liquid
reflecting on the surface after impact, as proposed in the skating droplet
model. However, while the splashing parameter $\beta$ from the dewetting theory is
not found to have predictive value for our simulations, the results do support
the hypothesis that a lifting force acting on the liquid sheet determines whether
it will break up or not.

Overall, the different models all provide elements of what is observed in the
simulations. The skating droplet model correctly predicts the existence and
scaling of a gas film under the droplet and the effect of pressure on the gas film
can possibly be captured by using a slip length to model the breakdown of the
continuum assumption. The lamella formation theory is able to
correctly predict the scaling of the lamella ejection velocity as function of the
impact velocity for various liquids, and lastly, the dewetting theory's
hypothesis of a lift force acting on the liquid sheet after ejection is
consistent with our results.

\begin{acknowledgments}
The authors would like to thank Sidney R. Nagel, Michelle Driscoll, Casey
Stevens, Irmgard Bischofberger, Michael Brenner, and Shmuel Rubinstein for many
fruitful discussions of the simulation results.  
\end{acknowledgments}

%

\end{document}